\documentclass[%
 reprint,
superscriptaddress,
 amsmath,amssymb,
 aps,
prb,
]{revtex4-1}

\usepackage{hyperref}

\hypersetup{}

\usepackage{units}
\usepackage[usenames, dvipsnames]{color}
\usepackage{graphicx}
\usepackage{dcolumn}
\usepackage{bm}
\usepackage[english=nohyphenation]{hyphsubst}
\usepackage[english]{babel}

\begin{document}


\title{Bistability in a Mesoscopic Josephson Junction Array Resonator}

\author{P.R. Muppalla}

\author{O. Gargiulo}
\affiliation{Institute for Quantum Optics and Quantum Information, A-6020 Innsbruck, Austria}
\affiliation{Institute for Experimental Physics, University of Innsbruck, A-6020 Innsbruck, Austria}

\author{S.I. Mirzaei}
\affiliation{Institute for Quantum Optics and Quantum Information, A-6020 Innsbruck, Austria}
\affiliation{Institute for Experimental Physics, University of Innsbruck, A-6020 Innsbruck, Austria}
\affiliation{Sharif University of Technology, Tehran, Iran}

\author{B. Prasanna Venkatesh}
\affiliation{Institute for Quantum Optics and Quantum Information, A-6020 Innsbruck, Austria}
\affiliation{Institute for Theoretical Physics, University of Innsbruck, A-6020 Innsbruck, Austria.}

\author{M.L. Juan}
\affiliation{Institute for Quantum Optics and Quantum Information, A-6020 Innsbruck, Austria}

\author{L. Gr\"unhaupt}
\affiliation{Physikalisches Institut, Karlsruhe Institute of Technology, 76131 Karlsruhe, Germany}

\author{I.M. Pop}
\affiliation{Physikalisches Institut, Karlsruhe Institute of Technology, 76131 Karlsruhe, Germany}

\author{G. Kirchmair}
\email{gerhard.kirchmair@uibk.ac.at}
\affiliation{Institute for Quantum Optics and Quantum Information, A-6020 Innsbruck, Austria}
\affiliation{Institute for Experimental Physics, University of Innsbruck, A-6020 Innsbruck, Austria}

\date{\today}

\begin{abstract}

We present an experimental investigation of stochastic switching of a bistable Josephson junctions array resonator with a resonance frequency in the GHz range. As the device is in the regime where the anharmonicity is on the order of the linewidth, the bistability appears for a pump strength of only a few photons. We measure the dynamics of the bistability by continuously observing the jumps between the two metastable states, which occur with a rate ranging from a few Hz down to a few mHz. The switching rate strongly depends on the pump strength, readout strength and the temperature, following Kramer's law. The interplay between nonlinearity and coupling, in this little explored regime, could provide a new resource for nondemolition measurements, single photon switches or even elements for autonomous quantum error correction.

\end{abstract}

\pacs{Valid PACS appear here}
\maketitle

The non-linearity provided by atoms and Josephson junctions is a necessary ingredient to observe quantum mechanical effects in cavity quantum-electro-dynamics (QED) and circuit QED (cQED) systems. Strong nonlinearites, much larger than the linewidth of the transition, are required to realize qubits~\cite{paik2011}, implement quantum information protocols~\cite{kelly2015,corcoles2015} and realize textbook quantum optics experiments~\cite{sun2014,wang2016}. Non-linearities much smaller than the linewidth of the transition are typically exploited for parametric processes~\cite{siddiqi2004,bergeal2010,roy2016} like amplification or frequency conversion at the quantum level.

Besides quantum information applications, there has been a growing interest to exploit cavity QED for ultralow-power classical logic elements~\cite{mabuchi2009,kerckhoff2010,bose2012}. This interest has been sparked by the ever growing all optical communication networks. Remarkably, a single photon transistor~\cite{chen2013}, reminiscent of an electronic transistor, has been implemented for the optical domain. In this device a single photon can switch a large optical field. Realizing such devices has been a challenging endeavour as the required non-linearity is hard to realize, due to the weak interaction of optical light with atoms. 

Much stronger light matter interactions can be achieved in the microwave regime using the cQED platform. In this context Josephson junction arrays (JJAs) have proven to be an ideal circuit element to build superconducting qubits with excellent coherence properties and unique tuning capabilities~\cite{manucharyan2009,pop2014,bell2016}. Similarly, JJAs have also been used to build quantum limited parametric amplifiers~\cite{macklin2015,castellanos-beltran2008,roy2016,eichler2014}. Recently, the coherence properties of the self resonances of JJAs~\cite{hutter2011,masluk2012}, as well as their self-Kerr and cross-Kerr coefficients have been measured~\cite{weissl2015}. The measured Kerr coefficient showed good agreement with a model based on a second order expansion of the Josephson potential~\cite{bourassa2012}. 

A regime of particular interest arises when the self-Kerr $K_i$ and cross-Kerr $K_{ij}$ nonlinear coefficients are on the order of the linewidth $\kappa$ of the system. In this regime the system will show a pronounced bistability~\cite{marthaler2006,guo2011} at the single to few photon level. Bistability is a phenomenon which is relevant in many fields, ranging from chemistry~\cite{garcia-muller2008} and biology ~\cite{mcdonnell2011,eguiluz2000} to Josephson junction physics~\cite{silvestrini1988,turlot1989} and cQED~\cite{mavrogordatos2017}. Very recently, an optically levitated nanoparticle has been shown to exhibit a stochastic bistability~\cite{ricci2017} and Kramers turnover~\cite{rondin2017}. 

In this Letter, we report on the realization of a JJA resonator with multiple modes and strong self-Kerr and cross-Kerr coefficients in the regime $K_i, K_{ij}\approx\kappa$. We investigate the bistability of one mode of the JJA and characterize the dependence of the switching rate on the pump strength, readout strength and temperature. 

\begin{figure}[t]
   \centering   
    \includegraphics{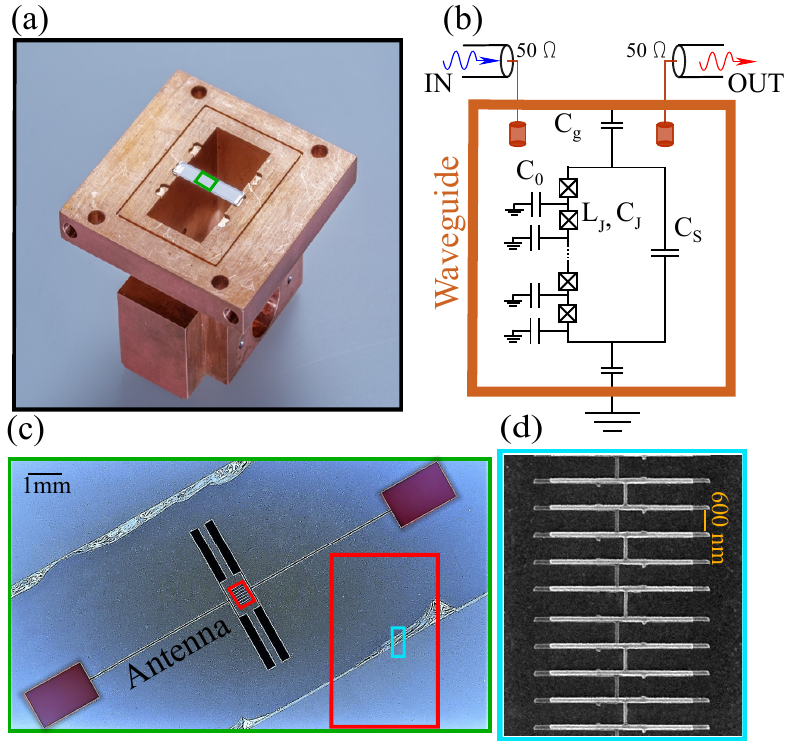}
    \caption{\textbf{(a)} Photograph of one half of a rectangular copper waveguide with a 6 GHz cutoff. A JJA with a microwave antenna is fabricated on a piece of sapphire and placed in the center of the waveguide \textbf{(b)} Schematic representation of an array of Josephson junctions inside a waveguide. $C_0$ is the capacitance of the islands to ground, $C_{S}$ is the shunt capacitance for the array, $C_J$ is the junction capacitance and $C_{g}$ denotes the coupling capacitance of the antenna to the waveguide.  $L_J$  is the Josephson junction inductance. The input and output couplers to the waveguide are shown on the top left and top right of the schematic. \textbf{(c)} Optical image of the JJA coupled to a 6~mm long antenna and a shunt capacitance. The inset shows a zoom-in on the  junction array. \textbf{(d)} Electron beam image (blue box in the inset) of ten of the $10^3$  Josephson junctions.}
    \label{Mfig1}
\end{figure}

The JJA consists of $10^3$ cascaded Josephson junctions, with a small capacitance to ground $C_0$, coupled to a 6~mm long microwave antenna and a shunt capacitance $C_s$, as shown in Fig.~\ref{Mfig1}. The junctions are fabricated on a sapphire substrate using electron beam lithography and bridge-free double-angle evaporation \cite{lecocq2011}. An electron beam image of the junctions can be found in  Fig. \ref{Mfig1}d. The junctions were designed to have a large ratio ${E_J}/{E_C} \approx 200 $, in order to suppress coherent quantum phase slips (CQPS)~\cite{pop2010,manucharyan2012}. Here $E_J$ is the Josephson junction energy and $E_C$ is the charging energy. The parameters of the JJA were designed such that the fundamental resonance of the JJA combined with the shunt capacitance is around 1~GHz. The mode spacing for the first 10 modes is about 1.2~GHz and progressively becomes smaller for higher resonances~\cite{masluk2012,weissl2015}.

The JJA is placed inside a  copper waveguide~\cite{kou2017a} with a 6~GHz cutoff, as shown in Fig.~\ref{Mfig1}a. Due to the capacitive coupling of the JJA to the waveguide, we can characterize the sample by performing microwave transmission measurements using a vector network analyzer (VNA). Due to the relative symmetry of the electric field of the waveguide and the antenna, the even modes of the JJA will couple poorly to the waveguide and not be visible in transmission measurements.
The waveguide with the sample is mounted on the mixing chamber stage (10~mK) of a cryogen free dilution refrigerator. The sample is enclosed in a double layer cryoperm shield inside a completely closed copper can. The stainless steel input lines are attenuated with 20~dB at 4~K and 30~dB at base temperature. They are  filtered with a combination of a 12~GHz low pass and an Eccosorb filter. The output stage consists of a 12~GHz low pass filter, two 4-12~GHz isolators and a 4-8~GHz high electron mobility transistor amplifier. The effective measurement bandwidth for direct transmission measurements using a VNA is limited to about 4-9~GHz due to the cutoff of the waveguide and the combined bandwidth of other microwave components.

Within the accessible measurement bandwidth we can characterize three resonances by fitting their transmission data to a notch type response function~\cite{probst2015a}. From these measurements we can extract the resonance frequencies ($\omega_i/2\pi$), the internal and the coupling quality factors (see supplementary material). 

 To observe the resonances of the modes outside the measurement bandwidth, we exploit the cross-Kerr interaction, which is induced by the junction non-linearity. The Hamiltonian for the JJA, up to second order, is given by
  \begin{equation}
    H/\hbar=\sum_{i=1}^N(\omega_i a_i^\dag a_i+\frac{K_{i}}{2}a_i^\dag a_ia_i^\dag a_i)+\sum_{\substack{i,j=1 \\ i \neq j}}^N K_{i,j}a_i^\dag a_ia_j^\dag a_j.
\end{equation}
The Hamiltonian consists of a self-Kerr term $K_i$ which leads to a photon number $n_i=a_i^\dag a_i$ dependent frequency shift of mode $i$ and a cross-Kerr interaction $K_{ij}$, which leads to a frequency shift of mode $i$ depending on the photon number in all other modes $j$. 

\begin{figure}[t]
    \centering
    \includegraphics{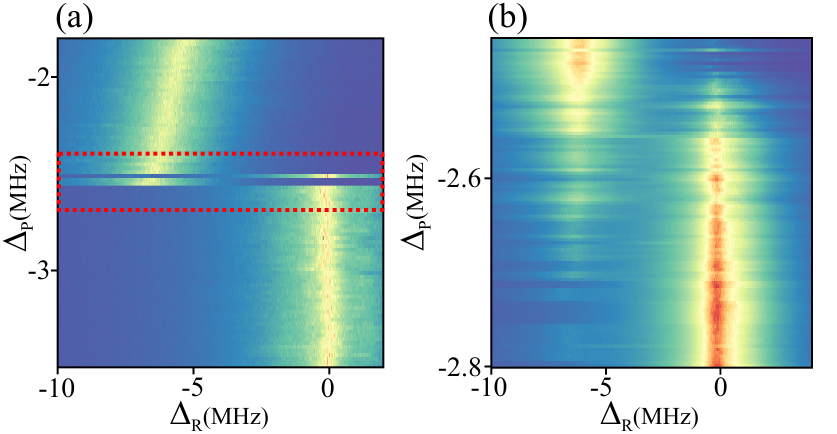}
    \caption{Two-tone measurement. \textbf{(a)} Shift $\Delta_R$ of the resonance frequency of the readout mode $\omega_{R}/2\pi =$ 7.105 GHz upon application of a pump tone. The pump tone is detuned from the resonance $\omega_{P}/2\pi =$ 4.8156 GHz of the 5th mode of the JJA by a detuning $\Delta_P$ and has a pump strength of $\bar{n}_{P}= 115$ photons. Due to the nonlinearity of the JJA, the resonance frequency of the readout mode is shifted when the pump tone matches mode five. \textbf{(b)} Zoom in on the bistable region of mode five. For a detuning of about $\Delta_P=-2.58(4)$~MHz two different resonance frequencies of the readout mode can be observed. The shifted and unshifted resonances correspond to 115 and 1 circulating photons in mode five, respectively.}
    
    \label{Mfig2}
\end{figure}

In Fig.\ref{Mfig2} we show the results of a two-tone spectroscopy where we sweep the frequency of a pump tone around mode five while weakly probing mode seven with the VNA. When the pump tone is resonant with mode five, we see a shift of the resonance frequency of mode seven. The measured frequency of mode five, using two-tone spectroscopy, matches the direct VNA measurement. We repeat this procedure, using mode seven as a readout, for frequencies between 900~MHz and 16~GHz. By comparing the measured frequencies of the entire JJA with the re-normalized mode frequencies calculated by diagonalizing the capacitance matrix (see supplementary material), we extract the JJA parameters $C_0$ = 0.152~fF, $C_J$ = 34~fF, $L_J$ = 1.25~nH, $C_{S}$ = 18~fF with a confidence range of about 20~\%. These parameters match well to the expected design values and room temperature resistance measurements of the junctions. 

\begin{figure*}[!t]
    \centering
    \includegraphics[width=\textwidth]{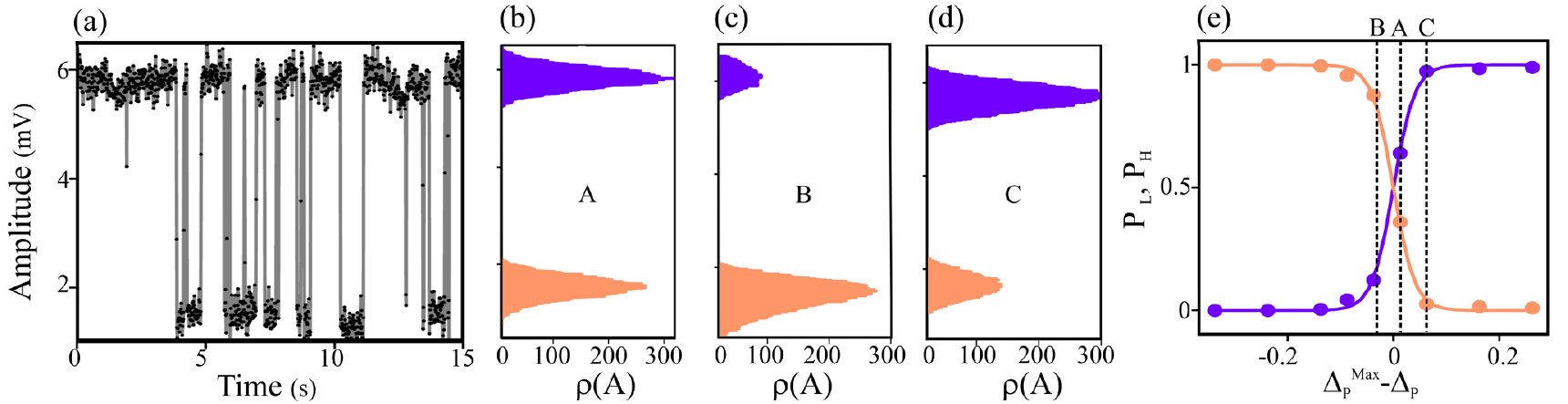}
    \caption{\textbf{(a)} Transitions from the low amplitude state to the high amplitude state for $\bar{n}_{P} = 9~$~photons at a detuning of $\Delta_P \approx \Delta_P^{\mathrm{Max}} =$~-0.54(1)~MHz from mode five. The readout mode seven was driven with $\bar{n}_{R} = 0.5 $~photons on resonance. The trace displayed here is a 15~s segment out of a total recorded time of 20000~s. \textbf{(b)} Histogram of the amplitude distribution $\rho(A)$ for the data displayed in \textbf{a}. \textbf{(c)} Histogram of $\rho(A)$ for a detuning of $\Delta_P =$~-0.59(1)~MHz.\textbf{(d)} Histogram of the amplitude distribution $\rho(A)$ for a detuning of $\Delta_P = $~-0.49(1)~MHz. \textbf{(e)}. Dependence of the normalized state population in the high and low photon state on $\Delta_P$ relative to the bistable point. The solid line is a fit to a sigmoid function.}
    \label{Mfig3}
\end{figure*}

With these parameters we can calculate $K_i$ and $K_{i,j}$ similar to Ref.~\cite{weissl2015}. The results are summarized in the supplementary material for modes five, seven and nine. Using these coefficients we can calibrate the photon number by measuring the resonance frequency shift as a function of the applied power (see supplementary material). The attenuation extracted from these Kerr measurements matches within 4\% to the attenuation in the cryostat, determined by independent transmission measurements.  

Upon closer inspection of the two-tone scan in Fig.\ref{Mfig2}a one can observe a bistable region (see Fig.\ref{Mfig2}b) for a detuning of $\Delta_P=-2.58$~MHz from the bare resonance frequency of mode five. Around this frequency, two different cross-Kerr shifts of mode seven can be observed: A shift of $\Delta_R = -7.32$~MHz corresponding to a photon number of about 115 photons in mode five and a shift of 133~kHz corresponding to about one photon. In the two-tone scan we observed residence times exceeding ten seconds in either the high or low photon number states.

To precisely characterize the residence time, we implemented a readout scheme similar to the dispersive state detection of a superconducting qubit in a circuit QED architecture~\cite{blais2004a}. We monitor mode seven continuously on resonance with a readout power $\bar{n}_R$ of about half a photon such that we can not see an apparent shift or broadening of mode seven. When we pump mode five with $\bar{n}_P$ photons, we observe a change in the transmitted readout signal corresponding to jumps between the high and the low amplitude states. For each data point we average 500 measurements each lasting 19.9~$\mu$s to get a good signal to noise ratio. In addition, the analog to digital converter needs another 5~ms to transfer the data. Thus, it takes about 15~ms to acquire one data point.

In Fig.\ref{Mfig3}a we show a typical time trace for a measurement time of 15~s and $\bar{n}_P$=9~photons in mode five. One can clearly observe two distinct amplitudes in transmission corresponding to two distinct photon numbers in mode five. We define the switching rate $\Gamma$ as the inverse of the average time between two transitions from the low to the high power state. The transient time between these two states is much faster than the data acquisition rate. 

A histogram of the data shows a well separated bi-modal distribution for the two amplitude states.  Furthermore, we extract the residence times in the high amplitude state, the resulting histogram shows an exponential behaviour typical for a  Poissonian statisics (see supplementary material).When we sweep the detuning of the pump frequency $\Delta_P$ across the bistable point we see a change of the relative height of the peaks in the amplitude distribution $\rho(A)$, Fig.\ref{Mfig3}b-e. There is a $\Delta_P^{\mathrm{Max}}$ for a given pump strength where the heights of the peaks are equal as shown for example in Fig.\ref{Mfig3}b.

\begin{figure}[t]
    \centering
    \includegraphics{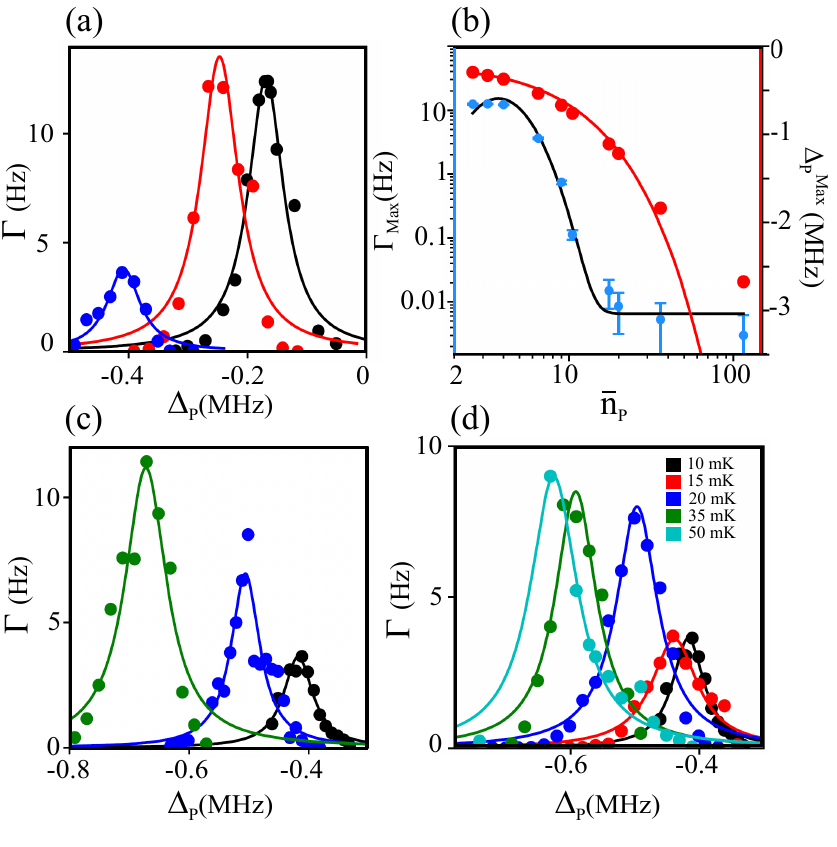}
    \caption{\textbf{(a)} Dependence of switching rate $\Gamma$ on $\Delta_P$ for three different pump strengths:  \textcolor{black}{\textbullet -}  $\bar{n}_{P}=2.6$~photons, \textcolor{red}{\textbullet -}  $\bar{n}_{P}=4$~photons, \textcolor{blue}{\textbullet -} $\bar{n}_{P}=$~6~photons. The solid lines are lorentzian fits to the data. \textbf{(b)} Extracted $\Gamma_{\mathrm{Max}}$ and corresponding $\Delta^{\mathrm{Max}}_{P}$ as a function of $\bar{n}_{P}$. The black line is a fit to our model (see text). \textbf{(c)} $\Gamma$ measured for constant $\bar{n}_{P}=6$~photons while varying the readout strength:  \textcolor{black}{\textbullet-}~$\bar{n}_R=0.5$~photons, \textcolor{blue}{\textbullet -}~$\bar{n}_R=$~1.5~photons, \textcolor{ForestGreen}{\textbullet -} $\bar{n}_R=$~2.5 photons. \textbf{(d)} $\Gamma$ measured for different cryostat base temperatures with $\bar{n}_P=6$. In (a), (b) and (d) $\bar{n}_R= 0.5$.}
    \label{Mfig4}
\end{figure}

We repeat the switching rate measurements for a total of ten different pump powers ranging from 2.6 to 115 photons in mode five. In Fig.\ref{Mfig4}b we plot the maximally achieved switching rate $\Gamma_{\mathrm{Max}}$ and the corresponding pump detuning $\Delta_P^{\mathrm{Max}}$ versus photon number. The lowest power we can observe switching for is $\bar{n}_P=2.6$~photons at a detuning of -169~kHz from the low power resonance. This detuning matches well to the prediction~\cite{ong2011} of $\Delta_P=\nicefrac{\sqrt{3}}{2}\, \kappa_5 =$~160~kHz with $\kappa_5 =~181$~kHz.

 Typically, stochastic switching in a bistable system is described by Kramers law~\cite{hanggi1990a}. There, the switching rate is determined by the potential landscape and the fluctuations. For a symmetric potential it is given by $\Gamma_{\mathrm{Max}} = \Gamma_0\, \mathrm{exp}(- E_b/k_b T)$, where $E_b$ is the barrier height between the two stable solutions and the prefactor $\Gamma_0$ which depends on the relative strength of the dissipation and the potential~\cite{hanggi1990a}. Here, the activation of the switching between the two stable solutions likely originates from the dispersive shift of the resonator frequency due to photon number fluctuations in the readout mode and thermal fluctuations of the photon number in all modes.

In our case, the potential landscape is created by the interplay between the pump, the self-Kerr effect and the damping of the mode. An intuitive choice for this potential~\cite{marthaler2006,guo2011} is provided by integrating the equation for the photon number in the steady state of a damped Kerr oscillator~\cite{agrawal1979}. From this model we extract the scaling of $E_b$~\cite{luchinsky1999} and $\Gamma_0$ for the maximum switching rate as a function of the pump photon number $\bar{n}_P$. We use this scaling in a fit function (see Fig.~\ref{Mfig4}b) with two free parameters to match our data (see supplementary material). We also add a $\Gamma_{\mathrm{res}}$ to take into account the finite switching rate for high pump powers. This finite rate could be limited by phase slips on the junctions of the JJA, which we estimated to be in the range of a few mHz. 

Additionally, we observe a change of the switching rate with respect to $\Delta_P$ following a lorentzian curve. The point of maximum switching rate $\Delta_P^{\mathrm{Max}}$ also corresponds to a symmetric amplitude distribution. In Fig.\ref{Mfig4}a one can see the change of the switching rate with respect to $\Delta_P$ for three different pump powers. The width of this lorentzian is about 72~kHz and the center shifts with increasing photon number in the pump. From our model we can also extract the shift of $\Delta_P^{\mathrm{Max}}$ with $\bar{n}_P$ which agrees well with the experiment as shown in Fig.~\ref{Mfig4}b. The deviation at high photon numbers can be explained by higher order Kerr effects which are not taken into account in the model.

To better identify the origin of the switching we also varied the power in the readout tone. The measured results are plotted in Fig.\ref{Mfig4}c for constant $\bar{n}_P = 6$~photons. We can observe two effects for an increasing photon number in the readout mode:  I) due to the cross-Kerr effect the bistable point moves to lower frequencies. II) in contrast to lowering the switching rate with the pump photon number, the readout photon number increases the switching rate. For $\bar{n}_R=2.5$~photons we see an increase in the switching rate by about a factor of three. This can be understood as a form of measurement induced dephasing. As the photon number in the readout resonator fluctuates, the position of the bistable point moves in frequency due to the cross-Kerr effect. This moves the pump in and out of the bistable region into the regime where either the low or the high power state are more likely, as the cross-Kerr frequency is on the order of the linewidth. This photon number fluctuation happens at a rate $\kappa_7=7.5$~MHz and is thus much faster than our acquisition time. For $\bar{n}_P >2.5$~photons, the switching becomes much faster and we cannot observe it any more due to the limited measurement bandwidth and signal to noise. Remarkably, changing $\bar{n}_R$ by only one photon for a constant $\Delta_P$ we can switch the state of mode five from the low to the high photon number occupation.

Lastly, to study the influence of thermal noise on the switching rate, we increased the cryostat base plate temperature from 10~mK to 50~mK ($\bar{n}_P=6$~photons, $\bar{n}_R = 0.5$~photons). Increasing the temperature increases the average thermal population as well as the fluctuations of the photon number in all of the modes of the JJA, most notably for the lower frequency modes. As a consequence, depicted in Fig.~\ref{Mfig4}d, we observe a shift in the magnitude and location of the maximum switching rate, similar to Fig.~\ref{Mfig4}c. This is again due to the cross-Kerr interactions of mode five with all other modes. The observed shift can be explained, by an increase in the JJA temperature up to about 100-130~mK depending on the initial temperature. We can get an upper bound for the JJA
temperature at the base temperature of the fridge, if we assume that the linewidth broadening we observe for mode five (see supplementary material) is due to the thermal population in the other modes of the JJA. This broadening can be explained by a minimal temperature of the JJA of $\approx 50$~mK. This is consistent with other cQED experiments~\cite{geerlings2013,jin2015}, where the devices are well above the fridge base temperature.

In conclusion, we have measured a stochastic bistability in a $10^3$ Josephson junction array which appears at a pump strength of only a few photons. This switching at low power is achieved by engineering the Kerr interaction strength to be comparable to the linewidth. We have seen an exponential decrease of the maximal switching rate for increasing pump strength as expected from Kramers law. For an increase in readout strength or temperature, the switching rate increases, likely due to photon induced dephasing through cross-Kerr interactions.

This proof of principle device demonstrates that it is possible for a few microwave photons in the readout mode to switch the photon occupation number by two orders of magnitude in the pump mode. As such it is a promising system to implement novel types of nondemolition measurements~\cite{kumar2010}, single photon microwave transistors~\cite{neumeier2013, manzoni2014}, single photon microwave switches~\cite{liao2009}, Flip-Flop memories~\cite{andersen2015} or even elements for autonomous quantum error correction~\cite{kerckhoff2010}.

We gratefully acknowledge fruitful discussions with Christian Kraglund Andersen, Alexandre Blais, Michel Devoret, Linghzen Guo, Archana Kamal, Zlatko Minev,  Michael Marthaler,  and Peter Talkner. We also thank Christian Schneider for valuable comments on the manuscript. We thank Luigi Frunzio for help with the fabrication of a first generation junction array resonator. IMP and LG acknowledge funding provided by the Alexander von Humboldt Foundation in the framework of the Sofja Kovalevskaja Award endowed by the German Federal Ministry of Education and Research. Facilities use was supported by the KIT Nanostructure Service Laboratory (NSL). MP, OG, PV, GK is supported by the Austrian Federal Ministry of Science, Research and Economy (BMWFW). MLJ is supported by European Council (EUP0252). 

%

\clearpage
\onecolumngrid
\begin{center}

\newcommand{\beginsupplement}{%
        \setcounter{table}{0}
        \renewcommand{\thetable}{S\arabic{table}}%
        \setcounter{figure}{0}
        \renewcommand{\thefigure}{S\arabic{figure}}%
     }

\textbf{\large Supplemental Material: Bistability in a Mesoscopic Josephson Junction Array Resonator }
\end{center}

\newcommand{\beginsupplement}{%
        \setcounter{table}{0}
        \renewcommand{\thetable}{S\arabic{table}}%
        \setcounter{figure}{0}
        \renewcommand{\thefigure}{S\arabic{figure}}%
     }

\setcounter{figure}{0}
\setcounter{table}{0}
\setcounter{page}{1}
\makeatletter
\renewcommand{\theequation}{S\arabic{equation}}
\renewcommand{\thefigure}{S\arabic{figure}}
\renewcommand{\bibnumfmt}[1]{[S#1]}
\renewcommand{\citenumfont}[1]{S#1}
\vspace{0.8 in}

\newcommand{\D}{\Delta}
\newcommand{\tD}{\tilde{\Delta}}
\newcommand{\K}{K_{PP}}
\newcommand{\bn}{\bar{n}_P}
\newcommand{\G}{\Gamma}
\newcommand{\LH}{\underset{L}{H}}
\newcommand{\HL}{\underset{H}{L}}
\newcommand{\blue}[1]{{\color{blue} {#1}}}

\section{Experimental setup and transmission measurements}

\begin{figure}[h]
    \centering
    \includegraphics{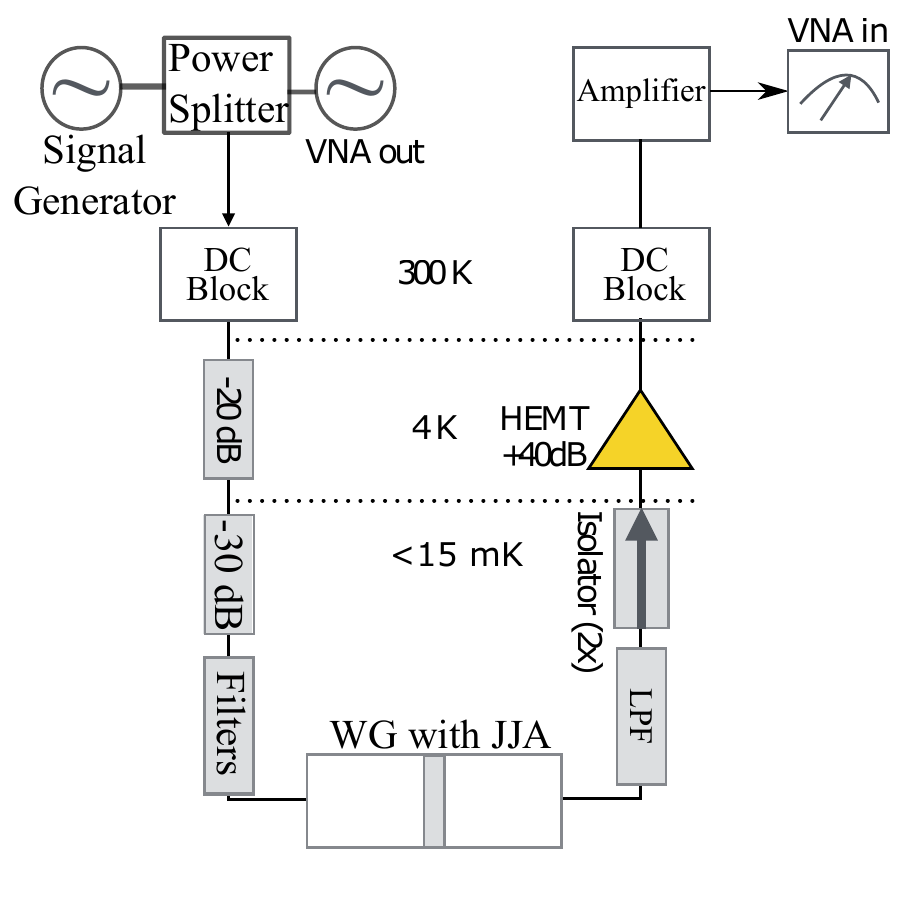}
    \caption{Experimental setup for performing two tone spectroscopies. For the switching rate measurements the VNA is replaced with a signal generator and mixers to down-convert the transmitted signal in order to digitize it with a 1~GS analog-to-digital converter.}
    
    \label{SFig.5}
\end{figure}

Fig.~\ref{SFig.5} shows the experimental apparatus used in the experiments. The waveguide with the sample is mounted on the mixing chamber stage (10~mK) of a cryogen free dilution refrigerator. The sample is enclosed in a double layer cryoperm shield inside a completely closed copper can. The stainless steel input lines are attenuated with 20~dB at 4~K and 30~dB at base temperature. They are  filtered with a combination of a 12~GHz low pass and Eccosorb filters. The output stage consists of a 12~GHz low pass filter, two 4-12~GHz isolators and a 4-8~GHz high electron mobility transistor (HEMT) amplifier. We additionally use DC-blocks at room temperature.

\begin{figure}[h]
    \centering
    \includegraphics{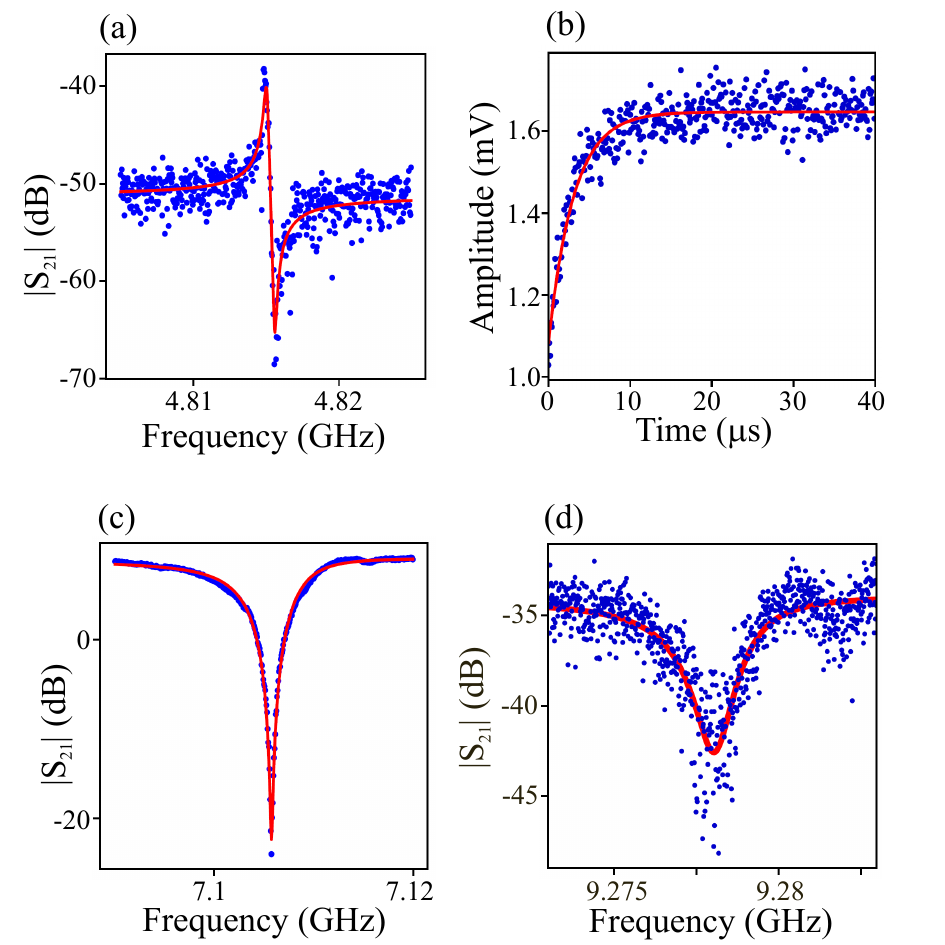}
    \caption{Transmission measurements, measured with a power corresponding to about one photon circulating in the resonator. The solid line in \textbf{(a), (c), (d)} is a fit using Eq.~\ref{eqn:1} to extract $Q_\text{tot}$ and $\omega_R$. \textbf{(a)} Resonator response measurement for mode five \textbf{(b)} Relaxation time $T_1$ on mode five with a pump strength of $\bar{n}_P = 115$ photons. The solid line is an exponential fit to the data with $T_1 = 3~\mu$s. \textbf{(c)} Resonator response measurement for mode seven. \textbf{(d)} Resonator response measurement for mode nine.}
    \label{SFig.6}
\end{figure}

Using a vector network analyzer (VNA), we measure the transmission $S_{21}$ of the waveguide. Since the sample is strongly capacitively coupled to the waveguide, we are able to directly detect some of the resonances on the network analyzer. Three resonant modes out of the 1000 resonances of the array are shown in Fig.~\ref{SFig.6}a, c, d. We fit the measured $S_{21}$ parameter using a notch type response function to determine the quality factors and the resonance frequencies:
\begin{equation}
    S_{21}=1-\frac{\frac{1}{Q_{ext}}-2\imath\frac{\delta f}{f_R}}{\frac{1}{Q_{tot}}+2\imath\frac{f-f_R}{f_R}}.
    \label{eqn:1}
\end{equation}
The extracted parameters are summarized in Table~\ref{table:1}.

 \begin{table}[h]
\centering
 \begin{tabular}{c c c c c c c c} 
 \hline
 Mode & $f_r $ & $Q_\text{tot}$  & $\kappa_i$& $K_{i}$ & $K_{i7}$\\
 & (GHz)&  & (kHz)& (kHz)&(kHz)\\
 \hline
 5 & 4.816(1) &26000&181 & 66 & 187 \\ 
 \hline
 7 & 7.1058(2) &950 & 7500& 133 & - \\
 \hline
 9& 9.278(1) & 3375 & 2750 & 218 & 343 &  \\
 \hline
\end{tabular}
\caption{Parameters for the three array modes that can be directly measured with the VNA. $f_r$ and $Q_\text{tot}$ were extracted from data. The Kerr and cross-Kerr coefficients are calculated from a fit to the dispersion relation Fig.~\ref{SFig.7}d.}
\label{table:1}
\end{table}

Fig.~\ref{SFig.6}b shows a measurement of the decay time $T_1$ of mode five for a pump strength of about $\bar{n}_P =$~115~photons. To excite the resonator the pump was detuned by about -2.54~MHz from the bare frequency $\omega_P/2\pi =$~4.8156~GHz of mode five. Similar to a two-tone measurement we use mode seven as a readout, with a readout strength of $\bar{n}_R =$~0.5~photons. To perform the decay time measurement we excite mode five for a few $\mu$s with a pulse before performing a readout on the pump mode again using a pulse of a few $\mu s$. By varying the delay between the two pulses we measure how the excitation decays over time. From the measurements we find $T_1 \approx 3 \mu s$. 
Comparing this to the direct VNA measurements we see that we have an additional dephasing mechanism which broadens the resonances. As explained in the main text we believe this is due to a fluctuating cross Kerr shift induced by the thermal population in the other modes.

\section{Two-tone measurements and Dispersion relation}

\begin{figure}[h!]
    \includegraphics[width=0.8\textwidth, height=12.5cm]{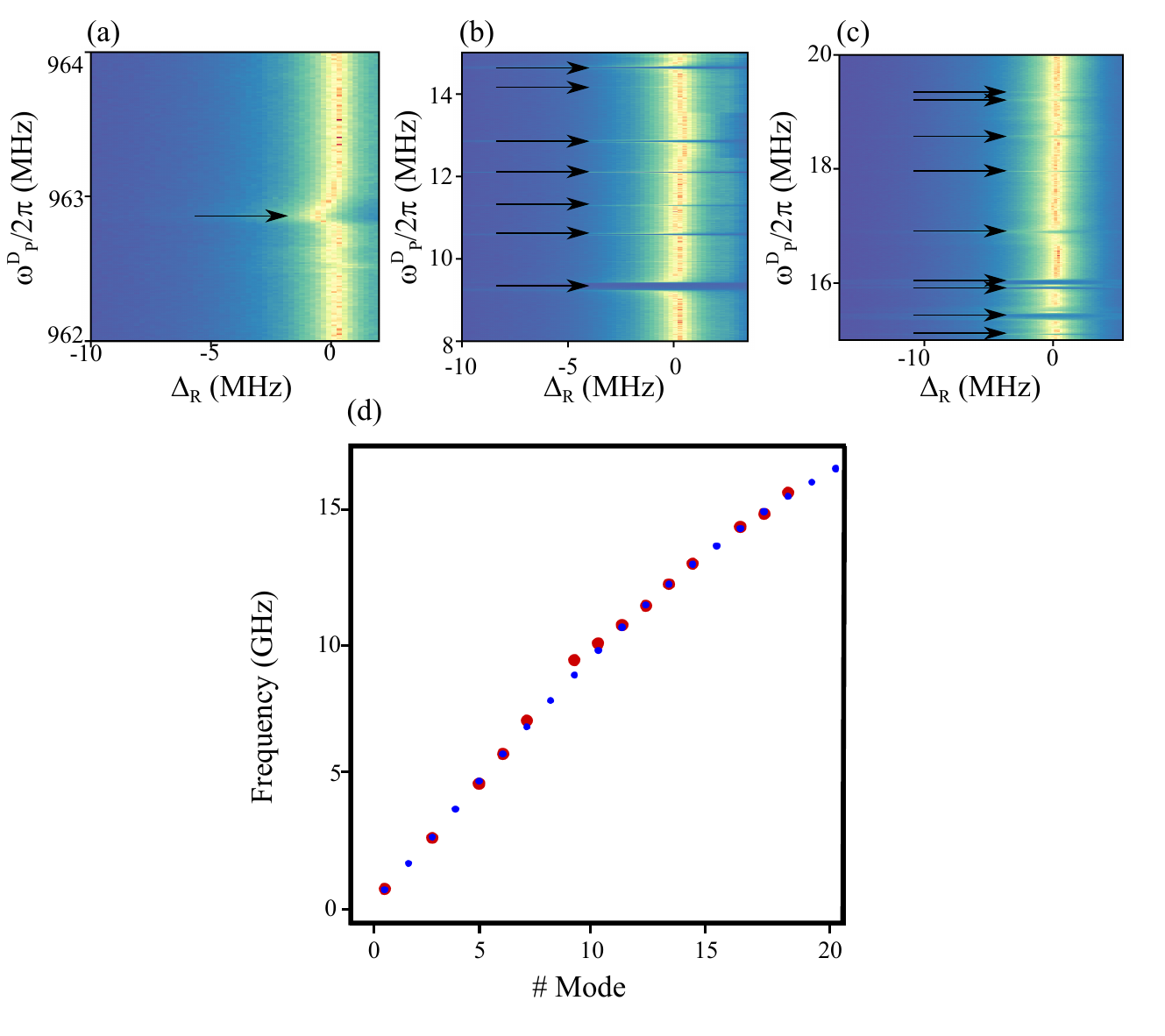}
    \caption{Two-tone measurements and dispersion curve. \textbf{(a)},\textbf{(b)} and \textbf{(c)} show the transmitted amplitude of the readout mode with frequency $\omega_R/2\pi =$ 7.105~GHz as a function of the pump tone frequency $\omega_P^D$ and detuning $\Delta_R$. The black arrows mark the frequencies when the pump tone matches a mode of the resonator, $\omega_P^D$, leading to frequency shift in the read out tone. \textbf{(d)} Measured resonant mode frequencies of the JJA from the two-tone spectroscopy. \textcolor{blue}{\textbullet} represents the calculated dispersion relation without including corrections due to the cross coupling between segments of the array, and \textcolor{red}{\textbullet} represents the measured resonant mode frequencies up to 20~GHz. }
   
    \label{SFig.7}
\end{figure}

Since the other resonance frequencies of the JJA's are out of the HEMT amplification bandwidth or below the cutoff frequency of the waveguide, we utilize a two-tone measurement technique to indirectly measure them. We use mode seven as read-out and apply a second pump tone. We sweep the frequency of this pump tone from ~960~MHz to ~20~GHz. Such a two-tone spectroscopy measurement can be seen as a pump probe experiment. The VNA is used to monitor one mode while a signal generator is used to excite a second one. We then observe the shift in the read-out mode due to populating other resonances through the pump tone. From this we can observe e.g. the fundamental mode of the array at ~963~MHz and other higher resonant modes up to ~20~GHz which is the limit of our signal generator. Two tone measurements for our array are shown in  Fig. ~\ref{SFig.7}. 

In Fig.~\ref{SFig.7}d, we show the dispersion relation of the array. The frequencies outside the HEMT bandwidth are extracted by two-tone spectroscopy measurements as shown in Fig.\ref{SFig.7}a, b, c. The mode spacing for the first 10 modes is about 1.2~GHz. For higher resonant modes the spacing between modes becomes smaller. The \textcolor{blue}{\textbullet} are obtained by diagonalizing the capacitance matrix which includes the shunt capacitance ($C_s$), junction capacitance ($C_J$), the ground capacitance ($C_0$) and the Josephson inductance ($L_J$) for the entire array structure. Most of the even modes can not be observed from two tone measurements as the electric field distribution of these modes has a symmetry that does not couple to the waveguide. For higher mode numbers this symmetry is somewhat broken due to inhomogeneities of the junctions and we can again excite these modes. 

We also note a discrepancy between the model and the measurements for modes 9 and 10. This is most likely due to the capacitive cross coupling between the parallel segments of the chain of Josephson junctions (see Fig.~1 in main manuscript). This effect can be accounted for by introducing additional capacitances with a minimal impact on the values of the self-Kerr and cross-Kerr coefficients for the lower modes.

\section{Kerr coefficients and Photon number calibration.}

The Kerr-nonlinearity  manifests itself as a frequency shift that depends on the circulating power in each of the resonant modes~\cite{Tweissl2015,Jbourassa2012}. By varying the input power on a low signal level, we are able to measure this frequency shift. The Self-Kerr and cross-Kerr shift can be used to calibrate the photon number in a resonant mode by measuring the frequency response as a function of drive power. 

\subsection{Self-Kerr Measurements.}

For the self-Kerr measurements only one mode $i$ of the chain is excited. We record the frequency shift $\Delta_P$ with respect to the input power using direct transmission measurements. We fit each data-set with a notch type response function to extract the frequency $f_r$ and $Q_\text{tot}$. From the shift relative to the bare resonance frequency we can extract the self-Kerr coefficient $K_i$.

\begin{figure}[h]
    \centering
    \includegraphics[width=\textwidth]{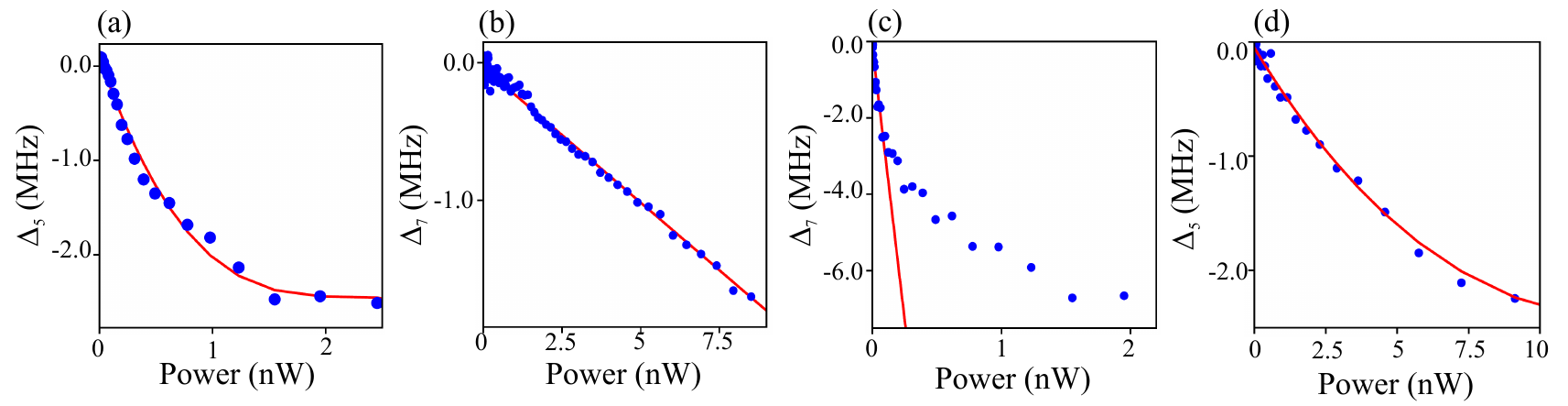}
    \caption{Self-Kerr and cross-Kerr Measurements. The error on each data point is about point size. \textbf{(a)} Dependence of the self-Kerr frequency shift on the input power of mode five. The red line is a third order polynomial fit - see text. \textbf{(b)} Dependence of the self-Kerr frequency shift on the input power of mode seven. The red line is a linear fit - see text. \textbf{(c)} Cross-Kerr frequency shift of mode seven when applying input power to mode five. The red line is calculated using the theoretical prediction for the linear self-Kerr term - see text. The read-out mode seven is driven with about $\bar{n}_R \approx 0.5$~photons.\ \textbf{(d)} Cross-Kerr frequency shift of mode five when applying input power to mode seven. The red line is a polynomial fit to the data. The read-out mode five is driven with about $\bar{n}_R \approx 1$~photons. }
    
    \label{SFig.8}
\end{figure}

Fig.~\ref{SFig.8}a shows the self-Kerr measurements on mode five. For low input power $P$ we observe that  $\Delta_5$ changes linearly with power. For high input power, higher order Kerr terms ($K_i', K_i''$) start to play a role. To take this into account we fit the following dependence 

\begin{equation}
\Delta_i(P)=K_i P+\frac{K_i'}{2}P^2+\frac{K_i''}{3}P^3 
\label{eq:2}
\end{equation}
to the data. With this we can extract the Kerr coefficients in units of Hz/W.

Fig.~\ref{SFig.8}b shows the self-Kerr measurements on mode seven. Here we observe that $\Delta_7$ changes linearly with the power due to the higher linewidth ($\kappa_7$) of this mode.

\subsection{Cross-Kerr Measurements $K_{75}$ utilizing two-tone spectroscopy}

For determining the cross-Kerr coefficients $K_{ij}$ we utilize two-tone spectroscopy measurements. One mode $i$ of the array is used as the readout with a power of about one photon and the other mode $j$ is excited with varying power. From the shift $\Delta_R$ of the resonance frequency of the readout mode $\omega_R/2\pi$ upon application of a pump tone, we can extract the cross-Kerr shift $K_{ij}$.  

\begin{figure}[h]
    \centering
    \includegraphics{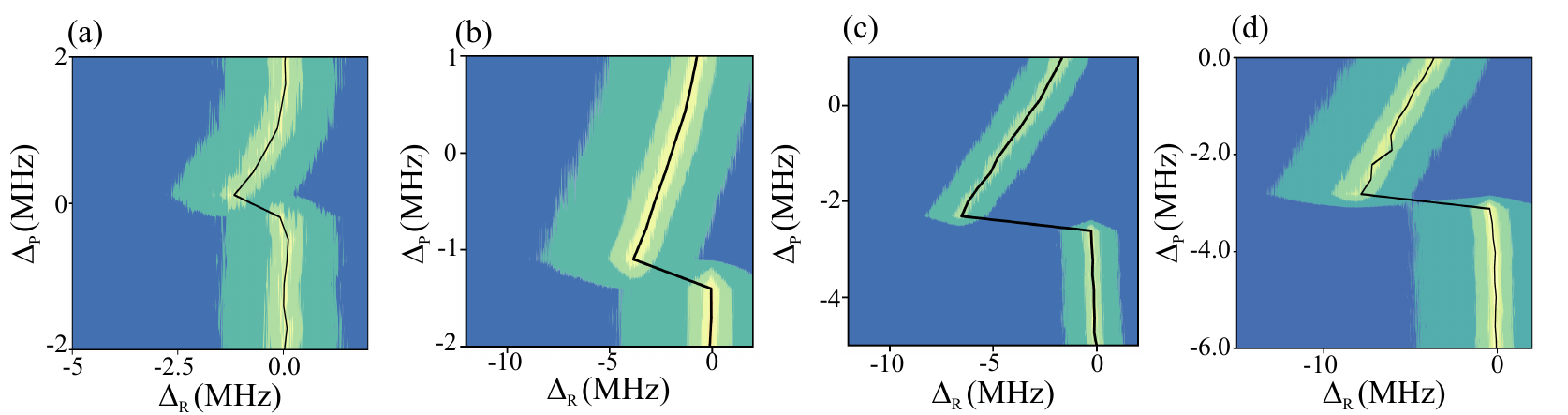}
    \caption{Cross-Kerr measurements $K_{75}$. \textbf{a, b, c, d} Two-tone spectroscopy measurements for different pump strength. Shift $\Delta_R$ of the resonance frequency of the readout mode $\omega_R/2\pi = 7.105$~GHz with $n_R = 0.5$~photons upon application of a pump tone to mode five. The pump frequency is detuned by $\Delta_P$ from the resonance of mode five $\omega_p = 4.1856$~GHz. \textbf{(a)},\textbf{(b)}, \textbf{(c)} and \textbf{(d)} Corresponds to a pump power of~$\approx$~0.01~nW, $\approx$~0.7~nW, $\approx$~1~nW and $\approx$~2~nW respectively; the solid black line corresponds to fit where we extract the resonance frequency of the readout mode for each pump frequency. From the fits we then extract the maximal frequency shift of the read-out mode for a given pump power in mode five.}
    \label{SFig.9}
\end{figure}

Fig.~\ref{SFig.9}a shows a typical two-tone spectroscopy measurements to extract the cross-Kerr coefficient $K_{75}$ by pumping mode five and observing the frequency shift on mode seven. We drive mode seven with a constant readout strength of $\bar{n}_R =$~0.5~photons. To extract the maximal frequency shift $\Delta_R$, we fit each pump frequency to a notch type response function and extract the resonance frequency(solid black line). The maximal shift of the readout resonator for a given pump power results in one datapoint in Fig.~\ref{SFig.8}c. We then repeat this procedure for different pump powers e.g. Fig.~\ref{SFig.9}b-c. From the measurement in Fig.~\ref{SFig.9} one can also clearly observe that the mode becomes bistable as the power is increased.

Fig.~\ref{SFig.8}c shows the result of all two tone measurement to determine the cross-Kerr shift $K_{75}$ when driving mode five and using mode seven as the readout. For low input power we observe that $\Delta_7$ changes linearly with the power but then rapidly higher order terms come into play. In this case, even a third order polynomial fit does not agree with the measured $K_{75}$. We choose instead to characterize cross-Kerr coefficient by the slope at low powers in Fig.~\ref{SFig.8}c. By using the dispersion relation fit in Fig.~\ref{SFig.7}d and the resulting diagonalized capicitance matrix, we extract the ratio between $K_5$ and $K_{75}$. This ratio, together with $K_5$ determined from the linear part of the fit function Fig.~\ref{SFig.8}a allows us to compute $K_{75}$. A linear fit with the computed slope $K_{75}$, shown by solid red line in Fig.\ref{SFig.8}c, shows good agreement with the measurements at low powers.

Fig.~\ref{SFig.8}d shows the cross-Kerr measurements $K_{57}$ on mode five using mode seven as the readout. Here the third order polynomial fit agrees well with the data. When we take the ratio of the first order self and cross-Kerr coefficients extracted from fitting the data in Fig.~\ref{SFig.8}b and Fig.~\ref{SFig.8}d and compare it to the Kerr coefficients extracted from fitting the dispersion relation we find an agreement within 99.8\%.

Furthermore, if we convert the Kerr coefficients extracted from Fig.~\ref{SFig.8}a-d from Hz/W to Hz/photon we find that the required conversion factor matches within 4\% to the attenuation in the cryostat, determined by independent transmission measurements.

For 0.1~nW input power we get about 10 photons in mode five. We can apply this same conversion factor to the whole polynomial Eq.~\ref{eq:2} and then use the observed frequency shift for a given input power to calculate the circulating photon number.

\subsection{Power dependence of the line-width}
\begin{figure}[h] 
    \centering
    \includegraphics{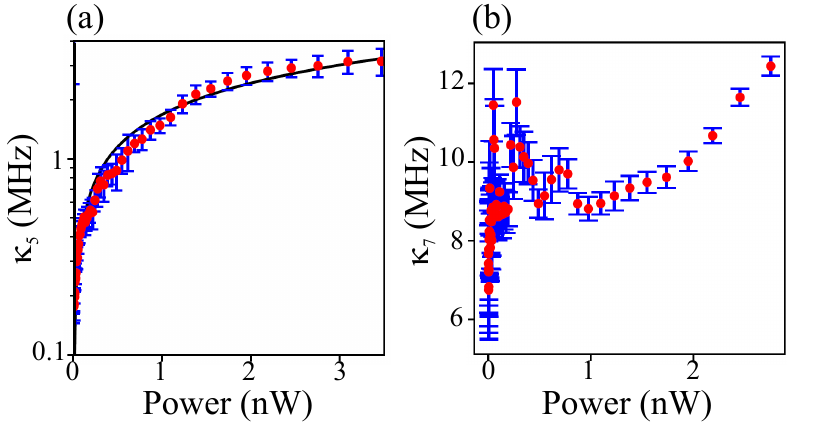}
    \caption{\textbf{(a)} Line-width $\kappa_5$ vs. input power on mode five. For the lowest drive power we still have a good enough signal to noise ratio to fit the data. We find $\kappa_{5} (\bar{n}_{P} \rightarrow 0) =$~181~kHz. The black solid line corresponds to a fit of a $\sqrt{\text{Power}}$ dependency.  \textbf{(b)} Line-width $\kappa_7$ vs. input power on mode seven. On the lowest drive power we still have a good enough signal to noise ratio to fit the data with a $\kappa_{7} (\bar{n}_{P} \rightarrow 0) = $~7.5~MHz. . }
    
    \label{SFig.10}
\end{figure}

Fig.~\ref{SFig.10} shows the linewidth of mode five and seven with respect to the input power. The line-widths are extracted from  direct transmission measurements. Each measurement is fit with a notch type response function to extract the resonance frequency $f_r$ and $Q_\text{tot}$ for a given drive power. The increase of the line-width $\kappa_i$ as a function of the circulating power is expected from the self-Kerr effect. For a circulating power of less than one photon in mode five we find $\kappa_5 (\bar{n}_{P} \rightarrow 0)  =$~181~kHz, and for mode seven $\kappa_7 (\bar{n}_{P} \rightarrow 0) =$~7.5~MHz.

\section{Width of bistable region for varying photon number}

\begin{figure}[h]
    \centering
    \includegraphics{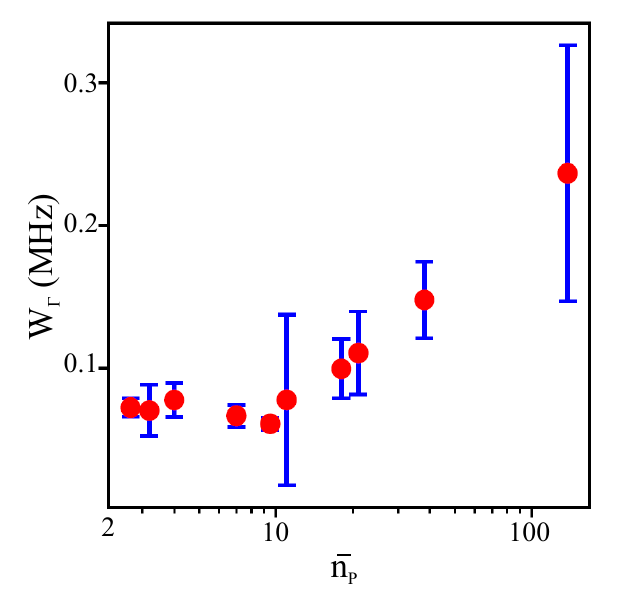}
    \caption{Width $W_\Gamma$ of the bistable region for different pump strengths $\bar{n}_p$.}
    \label{SFig.11}
\end{figure}

Fig.~\ref{SFig.11} shows the width of the bistable region obtained from the switching rate measurements (see main paper) for the ten different pump strengths shown in Fig.~4b in the main text. For very low pump strengths up to 10~photons the width of the bistable region is approximately constant $\approx$~72~kHz. For photon numbers greater than 10~photons the bistable region becomes wider.

\section{Residence time and state population inversion}

\begin{figure}[h]
    \centering
    \includegraphics{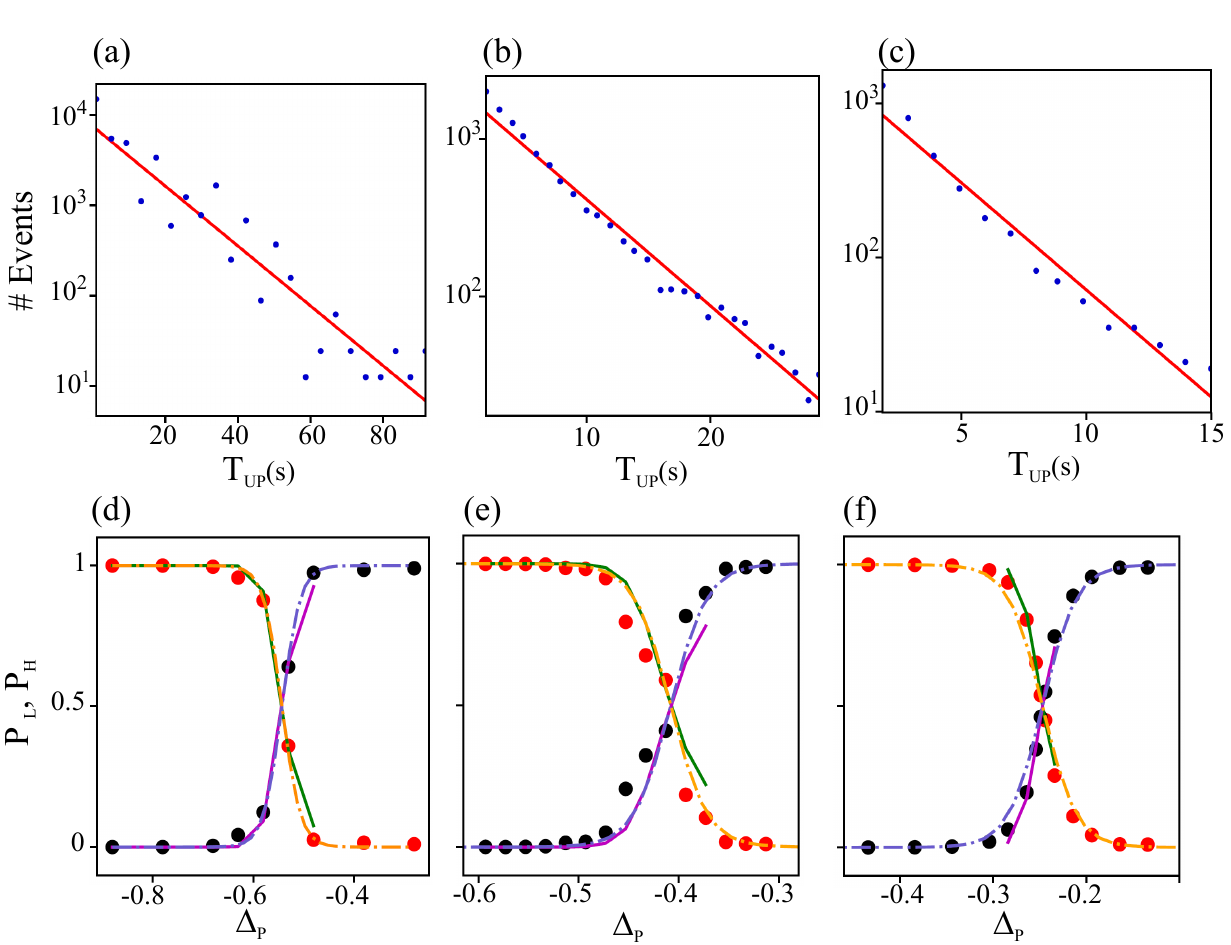}
    \caption{\textbf{(a)}. Histogram of the residence time $T_\mathrm{UP}$ for a detuning $\Delta_P^{\mathrm{Max}} =$~-0.54(1)~MHz from mode five in the high amplitude state for the time trace displayed in Fig.~3\textbf{a} (see main paper). The red line is an exponential fit to the data giving $\left < T_{UP}\right>=14.4$~s. \textbf{(b)}. Histogram of the residence time $T_\mathrm{UP}$ for a detuning $\Delta_P^{\mathrm{Max}} =$~-0.40(1)~MHz. The red line is an exponential fit to the data giving $\left < T_\mathrm{UP}\right> = 6.5$~s. \textbf{(c)}. Histogram of the residence time $T_{\mathrm{UP}}$ for a detuning $\Delta_P^{\mathrm{Max}} =$~-0.24(1)~MHz. The red line is an exponential fit to the data giving $\left < T_{\mathrm{UP}}\right> = 2.9$~s. \textbf{(d), (e), (f)} Probability of being either in the low-photon state or in the high-photon state in a bistable region. \textcolor{black}{\textbullet}, \textcolor{red}{\textbullet}  represents the measured data, \textcolor{Green}{\,\textbf{---}\,} , \textcolor{Plum}{\,\textbf{---}\,} are from fits using Eq. ~\ref{eq:8}, and \textcolor{Orange}{\,\textbf{---.}\,}, \textcolor{Violet}{\,\textbf{---.}\,} are fits to a sigmoid function. The pump strength is  $\bn = 9$ for $\textbf{(a), (d)}$, $\bn = 6$ for $\textbf{(b), (e)}$ and $\bn = 4$ for $\textbf{(c), (f)}$.}
    
    \label{SFig.12}
\end{figure}

Fig.~\ref{SFig.12}a, b, c show the exponential dependence of the residence time for a pump strength of $\bar{n}_{P} = 9~$~photons, $\bar{n}_{P} = 6~$~photons and $\bar{n}_{P} = 4~$~photons at a detuning of $\Delta_P = \Delta_P^{Max}$~MHz from mode five. This indicates that the transitions are random and follow a Poissonian statistics. From the exponential fit to the data we extract a mean residence time in the high state $\left < T_{UP}\right>$.

Fig.\ref{SFig.12}d, e, f show the probability for being either in the low or high photon state as we scan $\Delta_P$ across the bistable region for three different pump strengths $\bar{n}_{P} = 9~$~photons, $\bar{n}_{P} = 6~$~photons and $\bar{n}_{P} = 6~$~photons. It shows the state population inversion between the low photon and the high photon state, following a sigmoid behaviour $f(x)=1/(1+\textnormal{e}^{-x})$.

We note that the statistics of our results for different $\bn$, for both the residence time and the probability distributions, are consistent with bistable systems well described by the Kramers model~\cite{SHaenggi90}.

\section{Theoretical Model for the Switching Rates}

\begin{figure}[h]
    \centering
    \includegraphics[width=15cm,height=7.5cm]{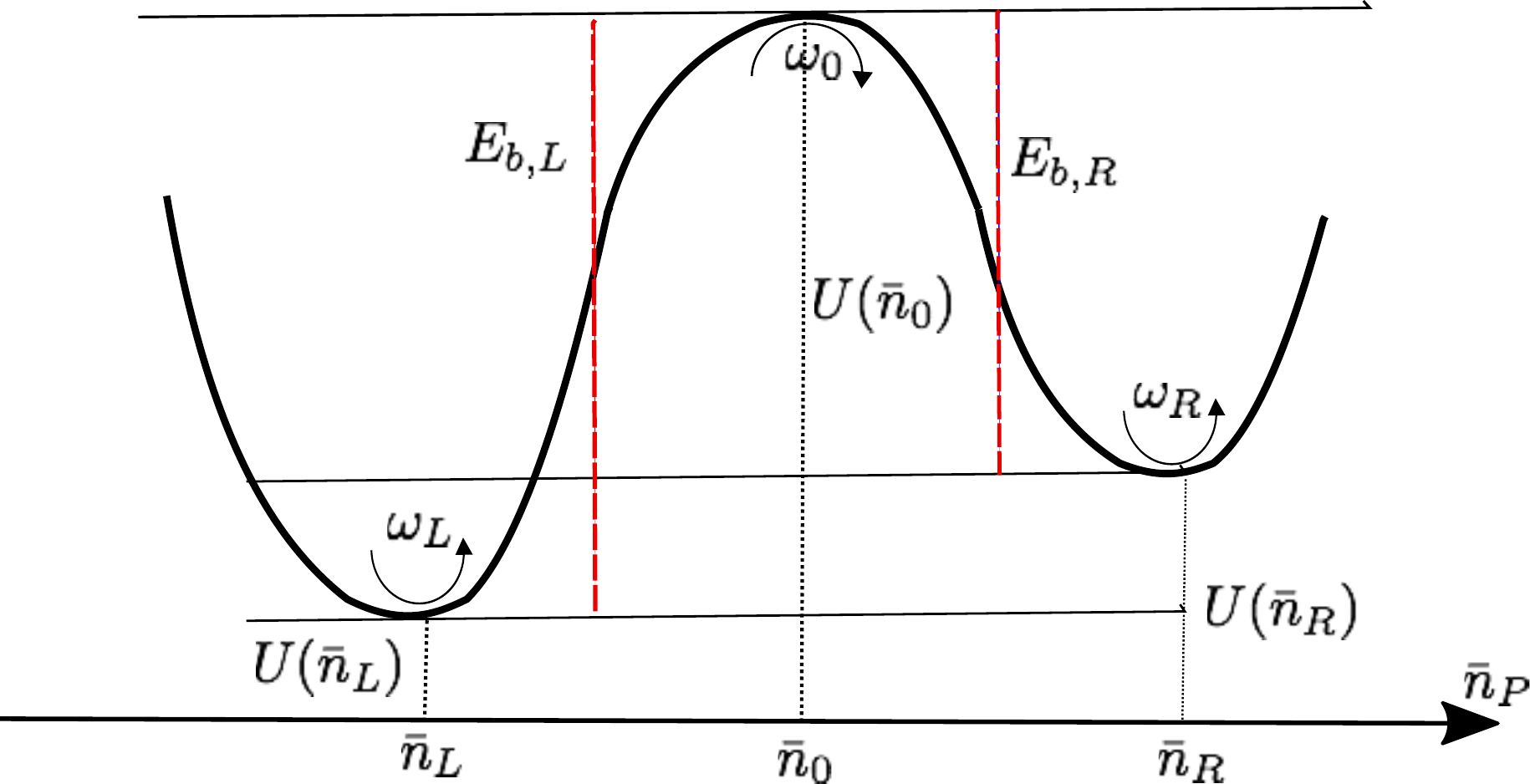}
     \caption{Schematic of the potential landscape for bistable switching.}
    \label{SFig.13}
\end{figure}

In this section we provide details regarding the theoretical model, based on Kramers theory of switching \cite{SHaenggi90}, used to obtain the fits in Fig.~4b of the main paper. To this end, we restrict to a simplified version of the full hamiltonian (1) of the main paper focusing only on the pump mode $5$ (indexed by $P$ in what follows) and include in addition a constant shift due to the cross-Kerr interaction $K_{PR}\bar{n}_R$ with the readout mode $7$ (indexed by $R$):
\begin{align}
H /\hbar = (\tD_P+K_{PR} \bar{n}_R) \hat{a}_P^{\dagger} \hat{a}_P+\frac{K_{PP}}{2} \left(\hat{a}_P^{\dagger} \hat{a}_P \right)^2+i \eta_P (\hat{a}^{\dagger}_P-\hat{a}_P) \label{eq:3}.
\end{align}
Here, we have written the hamiltonian in a rotating frame with respect to the driving of strength $\eta_P$ and with frequency $\omega_d$ and the detuning parameter $\tD_P = \omega_P - \omega_d$ with respect to the bare frequency of the pump mode $\omega_P$. Taking the classical limit of the Heisenberg-Langevin equation for the mode $\hat{a}_P$, we can obtain the following equation for the complex amplitude $\alpha_P = \langle \hat{a}_P \rangle$ mode (ignoring noise terms):
\begin{align}
\frac{d\alpha_P}{dt} = \left[-i(\tD_P+K_{PP}/2+K_{PR}\bar{n}_R) -i K_{PP} \vert \alpha_P \vert^2 -\frac{\kappa_P}{2}\right] \alpha_P + \eta_P 
\label{eq:4}
\end{align}
The steady state solution for this equation, defining $\delta = \tD_P+K_{PP}/2+K_{PR}\bar{n}_R$ written in terms of the average photon number $\bn = \vert \alpha_P \vert^2$ is:
\begin{align}
\bn = \frac{\eta_P^2}{\left[\delta + K_{PP} \bn \right]^2 + \kappa_P^2/4} 
\label{eq:5}   
\end{align}
This equation can have either one or three real roots. When it has three real roots, we have bistability. For a given value of $\delta$ and $K_{PP}$, it has three real solutions when $\vert \delta \vert > \sqrt{3} \kappa_P/2$ \cite{SAgarwal79,SOng11}, and the driving strength falls in the range $\eta_- \leq \eta_P \leq \eta_+$, where $\eta_{\pm}^2 = n_{c,\pm}\left(\left[\delta + K_{PP} n_{c,\pm}\right]^2 + \kappa_P^2/4 \right)$, with the extremal photon numbers given by
\begin{align}
n_{c,\pm} = \frac{-2 \delta}{3 K_{PP}} \left [1 \mp \sqrt{1-\frac{3}{4}\left(1+\frac{\kappa_P^2}{4\delta^2}\right)} \right] 
\label{eq:6}.
\end{align}
As we commented in the main paper, one key ingredient to analyze switching rates in a bistable system within the Kramers framework is a potential landscape in which the two stable solutions occur as local minima. A simple choice for such a potential is obtained by integrating Eq.~\ref{eq:5} \cite{SAgarwal79} with respect to $\bn$ (and dividing by $\eta_P^2$ to make it dimensionless) giving
\begin{align}
U(\bn) = \frac{K_{PP}^2}{4\eta_P^2}\bn^4 -\frac{2K_{PP} \delta}{3\eta_P^2} \bn^3+ \frac{1}{2\eta_P^2}\left(\delta^2+\kappa_P^2/4\right)\bn^2- \bn 
\label{eq:7}.
\end{align}
Note that this is not a real potential in the Hamiltonian sense, but a fictitious one for the average photon number $\bn$ treated as an independent degree of freedom. The critical points of $U(\bn)$, namely points where $dU/d \bn \equiv 0$ precisely satisfy Eq.~\ref{eq:5} and the solutions to the equations identify the extrema of the potential landscape. From the quartic form of the potential, as shown in Fig.~\ref{SFig.13}, we can anticipate that in the bistable region the potential has a double well shape with the two local minima at $\bn = \bar{n}_{L},\bar{n}_H$ and the local maxima (top of the barrier between the two wells) at $\bn = \bar{n}_0$ (with $\bar{n}_L<\bar{n}_0<\bar{n}_H$). Let us denote the oscillation frequencies at the bottom of the wells (top of the barrier) as $\omega_{L},\omega_H$ ($\omega_{0}$). The barrier height between $\bar{n}_{L}$ ($\bar{n}_H$) and $\bar{n}_{H}$ ($\bar{n}_L$) is given by $E_{b,L} = U(\bar{n}_0)-U(\bar{n}_L)$ ($E_{b,H} = U(\bar{n}_0)-U(\bar{n}_H)$). Given these parameters, the Kramers formula \cite{SHaenggi90} predicts that the rate of transition out of the wells are of the form $\G_{L\rightarrow H}=\G_{0,L\rightarrow H} \exp(-\beta_{eff} E_{b,L})$ ($\G_{H\rightarrow L}=\G_{0,H\rightarrow L} \exp(-\beta_{eff} E_{b,H})$) with the functional form of the pre-factor decided by the relative strengths of the damping rate and the well frequencies $\omega_{L,H},\omega_0$. In our treatment note that $\beta_{eff}$ is an effective dimensionless temperature. In addition in the limit of a two state model with localized states at $\bar{n}_{L}$ and $\bar{n}_H$, the average state population will be given by 
\begin{align}
P_{L} = \exp(-\beta_{eff} U(\bar{n}_L))/\left[\exp(-\beta_{eff} U(\bar{n}_L))+\exp(-\beta_{eff} U(\bar{n}_R))\right] 
\label{eq:8}
\end{align}
and $P_H = 1-P_L$. The total switching rate, $\G_{H\rightarrow L}+\G_{L\rightarrow H}$, is maximised when the energy barriers are the smallest and we find that this happens for a symmetric configuration with the same barrier height for both directions which we denote as $E_b$.

In order to compare the experimental results with the theoretical approach above, we first note that the parameter $\eta_P$ is not directly known but has to be inferred from the circulating photon number $\bn$. Let us consider the situation in Fig. 4b of the main paper, where the maximum switching rate and the associated detuning are plotted as a function of $\bn$ at the point of the maximum switch. In order to compare to the above model, by varying $\eta_P$ in the bistable region $\left[\eta_+,\eta_- \right]$ (with $K_{PP}$ and $\kappa_i$ chosen as the values in the experiment) we first numerically locate the detuning $\tD_P^{\mathrm{Max}}$ for a given value of $\bn$ satisfying the symmetric potential condition namely $U(\bar{n}_+)=U(\bar{n}_-)=U(\bar{n}_P)$. After adding a constant shift of $K_{PP}/2+K_{PR} \bar{n}_R$ to our numerically determined $\tD_P^{\mathrm{Max}}$, as shown by the solid red line in Fig. 4b of the main paper, we find good agreement with the experimental result. From Eq.\ref{eq:7}, we can extract the barrier height $E_b$, and frequencies $\omega_{L}$ and $\omega_0$ for this symmetric point as a function of $\bn$. In order to fit the switching rate to Kramers equation, we need to choose a form for the pre-factor $\G_0$. We found that in general the form $\G_0 \propto \omega_L \omega_0$ valid for the over-damped regime fits best to the experimental data. Fitting the measured rates $\G_{max}$ to the functional form $A \omega_L \omega_0 \exp(-\beta_{fit} E_b) + \G_{res}$ on a log-log scale using least squares procedure, we find the fit parameters $\G_{res} = (0.006 \pm 0.002)$ Hz, $A = (54.0 \pm 15.0)$ Hz and the effective temperature $\beta_{fit} = (2.4 \pm 0.2)$. The errors we have quoted here are the standard deviation on the estimated best fit parameters. The fitted curve was depicted by the black line in Fig. 4b in the main paper. We reiterate that in the fitting procedure the parameters $E_b$, $\omega_L$ and $\omega_0$ were calculated numerically from the potential Eq.~\ref{eq:7}. In addition for the cases of $\bn = 4,6,9$, we also fitted the form Eq. \ref{eq:8} (with $U(\bar{n}_{L,R})$ calculated numerically) for the average state population as a function of $\D_P$ and found the effective temperatures $\beta_{fit} = \{3.2 \pm 0.3,1.4 \pm 0.3,1.3 \pm 0.1\}$ respectively from the fits depicted in Fig. \ref{SFig.12} d, e, f. We can see that this temperature range is similar to the value we obtained from the fitting to Fig.4b. From Fig. \ref{SFig.12} d, e, f we can see that while the numerical model helps to locate $\D_P^{\mathrm{Max}}$ accurately, there are some discrepancies in terms of the width of the bistable region, especially regarding the $\Delta_P$ for the onset of bistability once we fix $\eta_P$.

\end{document}